\documentclass[sigconf, nonacm]{acmart}

\usepackage{csquotes}
\usepackage{pifont}
\usepackage{makecell}
\usepackage{listings}
\definecolor{lava}{rgb}{0.81, 0.06, 0.13}

\lstdefinelanguage{query}{
  sensitive=true,
  alsoletter={},
  alsoletter={},
  morekeywords={Initialize, Handle, Events},
  morecomment=[l]{//},
  morecomment=[s]{/*}{*/},
  morestring=[b]",
}

\newcommand{\codestyle}{\small\sffamily}

\usepackage{subcaption}
\usepackage{tikz}
\usetikzlibrary{arrows.meta, positioning, shapes.geometric, calc}
\usepackage{algorithm}
\usepackage[noend]{algpseudocode}
\usepackage{enumitem}
\usepackage{graphicx}
\usepackage[normalem]{ulem}

\newcommand{\holon}{\textsc{Holon Streaming}}

\newcommand{\ie}{\emph{i.e.,~}}

\makeatletter
\@ifundefined{showcomments}{
  \newcommand{\js}[1]{}
  \newcommand{\kk}[1]{}
  \newcommand{\rvg}[1]{}
  \newcommand{\ph}[1]{}
  \newcommand{\pc}[1]{}
}{
  \newcommand{\commentmacro}[3]{{\color{#1}$\triangleright$\emph{#2:~}{#3}$\triangleleft$}}
  \newcommand{\js}[1]{\commentmacro{blue}{JS}{#1}}
  \newcommand{\kk}[1]{\commentmacro{blue}{KK}{#1}}
  \newcommand{\rvg}[1]{\commentmacro{blue}{RVG}{#1}}
  \newcommand{\ph}[1]{\commentmacro{blue}{PH}{#1}}
  \newcommand{\pc}[1]{\commentmacro{blue}{PC}{#1}}
}
\makeatother

\begin{document}
\title{Holon Streaming: Global Aggregations with Windowed CRDTs}

\author{Jonas Spenger}
\orcid{0000-0002-7119-5234}
\affiliation{%
  \department{EECS and Digital Futures}
  \institution{KTH Royal Institute of Technology}
  \city{Stockholm}
  \state{Sweden}
}
\email{jspenger@kth.se}

\author{Kolya Krafeld}
\orcid{???}
\affiliation{%
  \department{EECS}
  \institution{KTH Royal Institute of Technology}
  \city{Stockholm}
  \state{Sweden}
}
\email{kolya@kth.se}

\author{Ruben van Gemeren}
\orcid{???}
\affiliation{%
  \department{EECS}
  \institution{KTH Royal Institute of Technology}
  \city{Stockholm}
  \state{Sweden}
}
\email{rubenvg@kth.se}

\author{Philipp Haller}
\orcid{0000-0002-2659-5271}
\affiliation{%
  \department{EECS and Digital Futures}
  \institution{KTH Royal Institute of Technology}
  \city{Stockholm}
  \state{Sweden}
}
\email{phaller@kth.se}

\author{Paris Carbone}
\orcid{0000-0002-9351-8508}
\affiliation{%
  \institution{RISE Research Institutes of Sweden}%
}
\affiliation{%
  \department{EECS and Digital Futures}
  \institution{KTH Royal Institute of Technology}
  \city{Stockholm}
  \state{Sweden}
}
\email{parisc@kth.se}

\begin{abstract}
  Scaling global aggregations is a challenge for exactly-once stream processing systems. Current systems implement these either by computing the aggregation in a single task instance, or by static aggregation trees, which limits scalability and may become a bottleneck. Moreover, the end-to-end latency is determined by the slowest path in the tree, and failures and reconfiguration cause large latency spikes due to the centralized coordination. Towards these issues, we present Holon Streaming, an exactly-once stream processing system for global aggregations. Its deterministic programming model uses windowed conflict-free replicated data types (Win\-do\-wed CRDTs), a novel abstraction for shared replicated state. Windowed CRDTs make computing global aggregations scalable. Furthermore, their guarantees such as determinism and convergence enable the design of efficient failure recovery algorithms by decentralized coordination. Our evaluation shows a 5x lower latency and 2x higher throughput than an existing stream processing system on global aggregation workloads, with an 11x latency reduction under failure scenarios. The paper demonstrates the effectiveness of decentralized coordination with determinism, and the utility of Windowed CRDTs for global aggregations.
\end{abstract}

\maketitle

\pagestyle{plain}
\begingroup\small\noindent\raggedright\textbf{Reference Format:}\\
\authors. \shorttitle.
arXiv preprint, 2025. 
\endgroup
\begingroup
\renewcommand\thefootnote{}\footnote{\noindent
  \textcopyright~The Author(s). This work is licensed under the arXiv.org perpetual, non-exclusive license 1.0. Visit \url{https://arxiv.org/licenses/nonexclusive-distrib/1.0/license.html} to view a copy of this license.
}\addtocounter{footnote}{-1}\endgroup



\section{Introduction}
Stream processing has become the de facto standard for real-time data processing~\cite{DBLP:journals/vldb/FragkoulisCKK24}.
Large enterprises use it to process billions of events per second across thousands of stream-processing pipelines \cite{mao2023streamops,fu2021real,DBLP:journals/pvldb/MeiLHLYXHCKW25}.
Aggregations over keys or partitions of the data, such as computing the most sold items in a category where each category is a key, are an important kind of computation in these pipe\-lines.
A \emph{global} aggregation, in contrast, is an aggregation that is computed \emph{across all} keyed partitioned streams.

Computing global aggregations in a traditional stream processing system~\cite{DBLP:journals/debu/CarboneKEMHT15,kreps2011kafka,sax2018streams,DBLP:journals/pvldb/AkidauBCCFLMMPS15}, however, has scalability issues.
In practice, the developer is forced to implement a global aggregation as a static \emph{aggregation tree}~\cite{flinkaggregations}, in which the aggregates are pre-aggregated at the leafs, combined by the nodes, and its final value computed at the root.
The root node, or any other node in this tree, may become a bottleneck for the computation.
The end-to-end latency is also a potential issue, as it is limited by the slowest path in the tree.
Moreover, failures and reconfiguration cause latency spikes when centralized coordination stops, reconfigures, and restarts.

We see this as an opportunity for using eventually consistent~\cite{DBLP:conf/sosp/TerryTPDSH95,brewertowards,vogels2009eventually} shared replicated state abstractions such as \emph{conflict-free replicated data types (CRDTs)}~\cite{DBLP:conf/sss/ShapiroPBZ11} within stream processing systems.
CRDTs are inherently decentralized and enable scalable computation of global aggregates through their algebraic properties such as commutativity, associativity, and idempotence, circumventing the issues as listed above.
However, they are not directly suitable for use in stream processing systems due to their weaker consistency model, making both their use and correct implementation in an exactly-once processing system cumbersome.

Towards this, we introduce \holon, a stream processing system with decentralized coordination for global aggregations.
\holon\ uses \emph{Windowed} CRDTs (WCRDTs), a novel abstraction introduced in this paper, to provide shared replicated state with stronger guarantees than regular CRDTs.
It builds on windowed computations~\cite{DBLP:conf/sigmod/LiMTPT05,DBLP:journals/pvldb/AkidauBCCFLMMPS15,DBLP:journals/vldb/VerwiebeGTM23} for slicing an infinite stream of data into an infinite sequence of finite windows.
Windowed CRDTs are guaranteed to provide deterministic output and converge on infinite streams.
This distinguishes them from regular CRDTs~\cite{DBLP:conf/sss/ShapiroPBZ11} which are not guaranteed to converge on an infinite stream;
a CRDT is never guaranteed to converge on infinite streams due to limitations of ``strong eventual consistency''~\cite{DBLP:conf/sss/ShapiroPBZ11}, \textls[-15]{it is only guaranteed to eventually converge ``if no new updates are\- made''~\cite{vogels2009eventually}.}
Moreover, \emph{all} programs written in \holon's programming model are guaranteed to be deterministic, enabled by its use of windowed CRDTs.
This includes programs consisting of WCRDTs and other forms of state, and their non-trivial composition such as branching on state.

Our implementation with decentralized coordination leverages the determinism and convergence of WCRDTs to provide a \emph{scalable} and \emph{low-latency} exactly-once processing system.
Global aggregations become scalable by means of leveraging the algebraic properties of CRDTs such as commutativity, associativity, and idempotence~\cite{DBLP:conf/sss/ShapiroPBZ11}.
The execution becomes low-latency by means decentralized coordination for execution, failure detection, failure recovery, and reconfiguration.
For this reason, \holon\ outperforms existing stream processing systems on global aggregations, with the potential to scale to unprecedented levels.
Our experiments on the Nexmark benchmark~\cite{tucker2002nexmark} demonstrate that \holon\ can achieve 2-times higher throughput and 5-times lower latency than when implementing the same workloads in Apache Flink~\cite{DBLP:journals/debu/CarboneKEMHT15}, with 11-times lower latency under failure scenarios.

\subsection*{Contributions}
In summary, this paper makes the following contributions.
\begin{itemize}
  \item We present \holon, an exactly-once stream processing system with decentralized coordination for global aggregations.
  \item We present its deterministic programming model based on \emph{Windowed CRDTs}.
    Windowed CRDTs are a novel abstraction for shared replicated state. Each partition is guaranteed to output the same window value.
    (\S\ref{sec:programming-model})
  \item We describe its architecture and implementation, providing algorithms for Windowed CRDTs, failure recovery, and reconfiguration by work stealing. (\S\ref{sec:system-overview})
  \item We evaluate \holon\ on global aggregation workloads and show that it outperforms existing stream processing systems on throughput and latency, and that it makes continuous progress even under failures and reconfiguration. (\S\ref{sec:evaluation})
\end{itemize}

\section{Background and Examples}\label{sec:background}

\subsection{Stream Processing and CRDTs}
\emph{Stream processing systems}~\cite{DBLP:journals/vldb/FragkoulisCKK24}, or stateful dataflow systems, are systems which process infinite streams of data in real time.
We are primarily concerned with exactly-once processing systems in this paper, such as Apache Flink~\cite{DBLP:journals/debu/CarboneKEMHT15}, Apache Kafka~\cite{kreps2011kafka,sax2018streams}, Timely Dataflow~\cite{murray2013naiad,murray2016incremental}, and Google Dataflow~\cite{DBLP:journals/pvldb/AkidauBCCFLMMPS15}.
These systems guarantee that each event is processed \emph{exactly once}, \ie that its side effects are visible exactly once in the state or output.

Perhaps surprisingly, none of these systems provide eventually consistent shared replicated state in the form of \emph{conflict-free replicated data types (CRDTs)}~\cite{,DBLP:conf/sss/ShapiroPBZ11}.
CRDTs are a class of data types with strong eventual consistency~\cite{DBLP:conf/sss/ShapiroPBZ11}, following work on replicated data types~\cite{shapiro2007designing,preguicca2009commutative} with eventual consistency~\cite{DBLP:conf/sosp/TerryTPDSH95,brewertowards,vogels2009eventually}.
That is, once all updates have stopped and the states have been synchronized, all replicas will converge to the same value.
Due to its weaker consistency guarantees, integrating CRDTs into a stream processing system is not trivial, as explored next.

\subsection{Global Aggregations}
\emph{Real-time global aggregations is an overlooked problem in stream processing systems}.
To compute a global aggregation, the user typically has to resort to writing a static aggregation tree~\cite{flinkaggregations}.
However, global aggregations have scalability issues.
Consider the following example query to illustrate the issues with existing systems.

\begin{quotation}
  \indent \indent \textit{Query 1:} Calculate the ratio of processed bids per partition relative to the global number of processed bids.
\end{quotation}

\begin{figure}[t]
  \centering
  \includegraphics[width=0.75\linewidth]{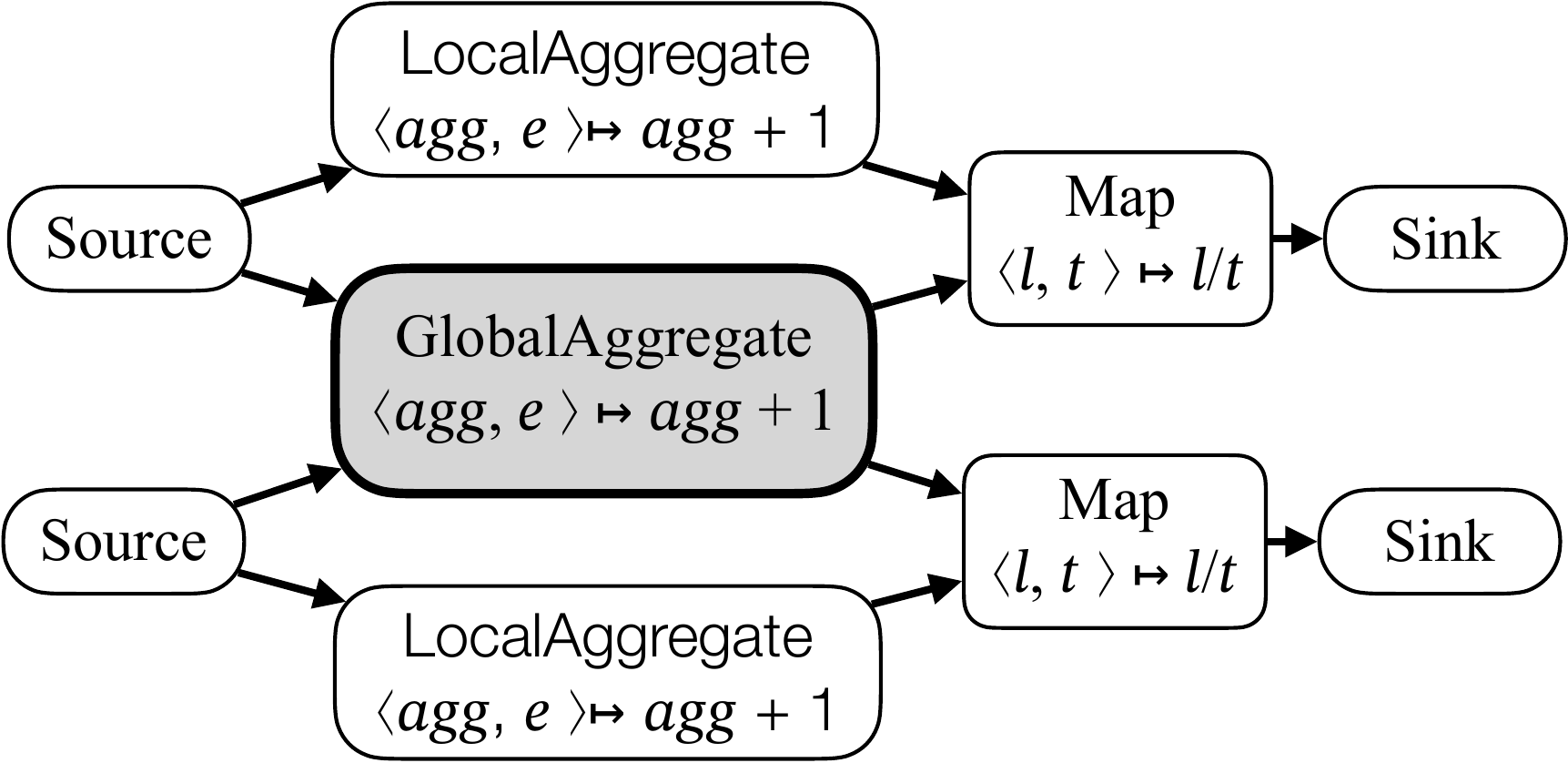}
  \caption{Query 1 physical dataflow execution plan.}\label{fig:dataflowquery}
\end{figure}

An implementation in a generic stream processing language is shown in \autoref{fig:dataflowquery}.
Here, we create a local aggregation per partition, and one global aggregation for all partitions.
The global aggregation will either be implemented itself as a single node instance, or by means of a tree~\cite{flinkaggregations,sparktreereduce}.

It is clear that the global aggregation may become a bottleneck for the computation.
This is especially the case for existing stream processing systems, as they typically have centralized coordination~\cite{DBLP:journals/pvldb/CarboneEFHRT17}.
That is, the entire pipeline shares some dependency, such that either all or none can complete the computation.

\begin{lstlisting}[language=query, caption={Query 1 with CRDTs.}, label={lst:query-crdts}, float={t}]
totalCount = GCounter { }
localCount = 0
for e in events do:
  if e is Bid then:
    totalCount.insert(1)
    localCount = localCount + 1
    emit $\langle$ localCount / totalCount.getValue() $\rangle$
\end{lstlisting}

Using CRDTs~\cite{DBLP:conf/sss/ShapiroPBZ11}, however, is non-trivial due to their weaker consistency guarantees.
Consider the CRDT-implementation shown in \autoref{lst:query-crdts}.
Here, the global aggregation is implemented as a Grow-only Counter (GCounter)~\cite{shapiro:inria-00555588}, and the local aggregation is a local variable.
The issue with this implementation is that the values which are read from the CRDT are nondeterministic, that is, if two nodes run \autoref{lst:query-crdts}, the calls to the method \lstinline{totalCount.getValue()} may return different values due to the nondeterministic network order.
Because of this, the computed ratio is nonsensical.

\subsection{Centralized vs Decentralized Coordination}
\emph{An inherent problem with existing stream processing systems is their centralized coordination.}
In an execution with centralized coordination, if a single node fails, then no new computations can be completed, and the entire system, consisting of all nodes, will eventually stop and restart~\cite{DBLP:journals/pvldb/CarboneEFHRT17}.
The same is true for network partitions as well as reconfigurations, causing the processing to stall until the system is reconfigured.
Furthermore, the computation only completes once all nodes have completed their work~\cite{DBLP:journals/pvldb/CarboneEFHRT17}.
For this reason, one node can cause the entire system to slow down.

In essence, the problem is related to the CAP theorem~\cite{brewertowards}.
A stream processing system cannot provide all three of consistency, availability, and partition tolerance.
Instead, however, it may be possible to use weaker forms of consistency such as eventual consistency and a more decentralized approach to overcome these issues~\cite{vogels2009eventually}.

\emph{Building a system with purely decentralized coordination, however, is not trivial.}
Strong eventual consistency~\cite{DBLP:conf/sss/ShapiroPBZ11} does not guarantee that the system will ever converge, due to the stream computation being infinite, it is only guaranteed to eventually converge ``if no new updates are made''~\cite{vogels2009eventually}.
Further, the eventual consistency model makes it difficult to reason about the program, as well as implementing failure recovery, as using CRDTs may lead to nondeterministic behavior.

\subsection{Determinism}
A global aggregation should ideally always return the same value for the same input for all participating nodes.
Determinism has two benefits:
(a)~it is easier to reason about deterministic programs,
(b)~it is easier to implement execution, failure recovery, and reconfiguration algorithms for deterministic programs.

In particular, deterministic programs always yield the same output and state for the same input, irrespective of execution and network order between the distributed nodes, which unlocks new failure recovery algorithms such as opportunistic work stealing.
For this reason, we are primarily interested in deterministic programming models.

\subsection{Goals and Non-Goals}
In light of these observations, we have chosen to design \holon\ by the following goals and non-goals.
The goals state important differences to existing systems, whereas the non-goals state deliberately omitted features for enabling the goals.

\textbf{Goals.}
\holon\ is a stream processing system with decentralized coordination, optimized for global aggregation workloads, with exactly-once processing guarantees, and a deterministic programming model.
\emph{Decentralized coordination is used to overcome the limitations of existing systems.}
That is, execution, failure recovery, and reconfiguration are local operations implemented by means of work stealing.
This is made possible by the use of CRDTs, and the lack of direct dependencies between partitions.
\emph{We have chosen the following CAP~\cite{brewertowards} trade-off for the system tailored for decentralized stream processing.}
Updating the state is highly available, enabling continuous progress even under network partition failures.
Reading the state is consistent, prioritizing deterministic reads of completed windows over immediate availability, as windows may need to wait for the advancement of the global watermark.
Importantly, determinism and decentralized coordination ensures that \emph{checkpointing and recovering state} is always available and consistent even under network partition failures.

\textbf{Non-goals.}
\emph{In contrast to existing systems, we deliberately exclude the following features to enable cleaner and stronger stream-processing semantics.}
No shuffles, \ie data is not shuffled between partitions, but state is asynchronously shuffled in the background for the CRDT synchronization.
No pipelining of execution nodes.
No consistent global state.
No centralized recovery.

\section{Programming Model}\label{sec:programming-model}
The \holon\ programming model is a stream processing model with Windowed CRDTs for shared state.\footnotemark~
Importantly, and for good reasons, it does not have a \emph{shuffle} operation, nor is its execution pipelined over separate computing nodes.
Instead, Windowed CRDTs are used to synchronize the state and communicate between the partitions.
These comparatively radical ideas make the implementation of \holon\ simple and efficient.
At the same time, its use is clear and straightforward for declaring global aggregations.

\footnotetext{
  The name \holon\ was chosen to reflect the compositional elements of the programming model and for the execution's decentralized nature.
  Parts are composed of parts, and the parts may be executed independently of each other.
  \begin{quote}
    ``A holon is something that is simultaneously a whole in and of itself, as well as a part of a larger whole. In this way, a holon can be considered a subsystem within a larger hierarchical system.''
    ~\cite{HolonWiki}
  \end{quote}
}

\subsection{Interface}
The \holon\ programming model has two APIs: a procedural API and a dataflow API.
This paper focuses on the the core interface of \holon\ which is its procedural API.
The dataflow API is implemented on top of the procedural API and provides a Flink~\cite{DBLP:journals/debu/CarboneKEMHT15}-like interface.

\begin{table}[t]
  \centering
  \caption{\holon's procedural API.}\label{tab:procapi}
  \vspace{-1.0em}
  \begin{tabular}{ll}
    \toprule
    \textbf{Operation} & \textbf{Description} \\
    \midrule
    \textbf{WCRDT interface} \\
    WCRDT \{ zero : Type \} & Create WCRDT of Type init zero. \\
    insert(element, ts) & Insert the element for timestamp ts. \\
    \quad$\rightarrow$ Unit\\
    getWindowValue(ts) & Read the winddow value for \\
    \quad$\rightarrow$ Option[Value] &  timestamp ts.\\
    incrementWatermark(ts) & Increment the local watermark to \\
    \quad$\rightarrow$ Unit & timestamp ts.\\
    getGlobalWatermark() & Read the global watermark. \\
    \quad$\rightarrow$ Timestamp\\
    \midrule
    \textbf{WLocal interface} \\
    WLocal \{ zero : Type \} & Create local windowed value. \\
    etc. \\
    \midrule
    \textbf{Local interface} \\
    Local \{ zero : Type \} & Create local value. \\
    etc. \\
    \bottomrule
  \end{tabular}
\end{table}

The procedural API consists of operations for creating Windowed CRDT values, Windowed Local values, and Local values (\autoref{tab:procapi}).
\emph{Local} refers to \emph{partition} local values.
A Windowed CRDT is created by calling the constructor and providing the type of the CRDT to create, as well as its initial value, and its windowing settings.
It has four methods for inserting elements, reading the window value, incrementing the watermark, and getting the global watermark.
There are two modes for reading a window value in the procedural API.
The safe mode blocks and awaits until the window value is completed before returning, and is safe to use for deterministic computations.
The unsafe mode returns immediately either with \lstinline{Some} value if the window is completed, or \lstinline{None} if the window is not yet completed.
All three state types are managed by the runtime system, that is, they are automatically checkpointed and recovered.

\subsection{Case Study: Query 1}

\begin{figure}
  \centering
  \includegraphics[width=1.0\linewidth]{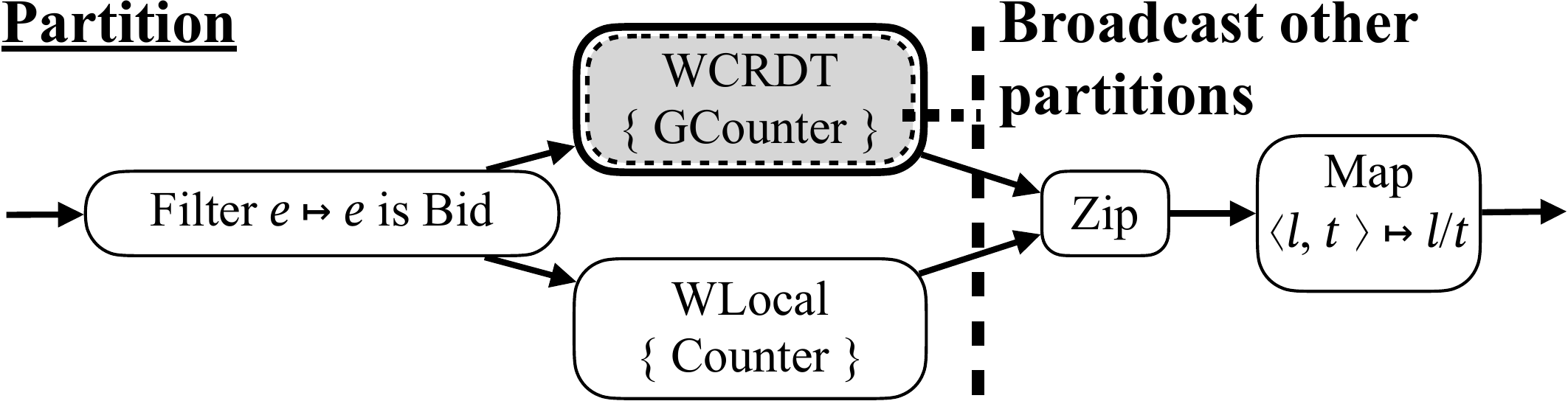}
  \caption{Query 1 in the dataflow API.}\label{fig:query1holon}
\end{figure}

To illustrate the programming model, consider \emph{Query 1} from \autoref{sec:background}.
\autoref{lst:query-proc} shows its implementation in the procedural API, together with a graphical representation of the dataflow execution plan in \autoref{fig:query1holon}.

\begin{lstlisting}[language=query, caption={Query 1 in the procedural API.}, label={lst:query-proc}, float={t}]
Initialize:
    totalCount = WCRDT { zero : GCounter }
    localCount = WLocal { zero : Counter }
    prevWatermark = Local { 0 : Integer }
Handle Events events:
    for e in events do:
        if e is Bid then:
            totalCount.insert(1, e.ts)
            totalCount.incrementWatermark(e.ts)
            localCount.insert(1, e.ts)
            localCount.incrementWatermark(e.ts)
    watermark = totalBids.globalWatermark()
    for w in prevWatermark ... watermark do:
        total = totalCount.getWindowValue(w)
        local = localCount.getWindowValue(w)
        emit $\langle$w, local / total$\rangle$
        prevWatermark = watermark
\end{lstlisting}

\begin{figure}[t]
  \centering
  \includegraphics[width=\linewidth]{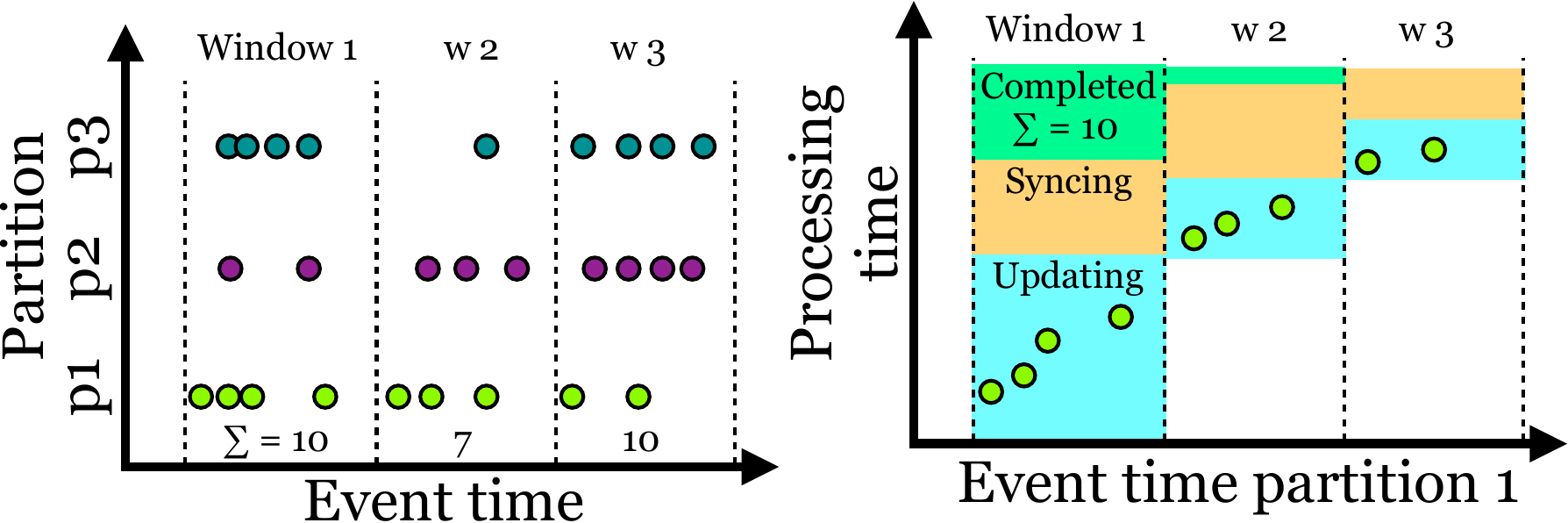}
  \vspace{-1em}
  \caption{Windowing updates for the GCounter.}
  \label{fig:windowing}
\end{figure}

To calculate the ratio of processed bids per partition, we choose to use a Windowed CRDT of the global count of bids, and a local windowed counter for the local count of bids.
These two windowed output values are zipped together, to calculate the ratio of local to global bids per window for each partition.
This is emitted to the output of the processing function.
\autoref{fig:windowing} illustrates the windowing semantics with tumbling windows for the GCounter.
In the first window, the GCounter is incremented ten times, the synchronization of a windowed CRDT, however, is not immediate, rather has three phases (updating, syncing, completed), after which it is available to read.
The listing assumes that the input events arrive in timestamp order per partition for simplification, it is possible to handle out-of-order events as well by calculating the watermark in a different way~\cite{akidau2021watermarks,DBLP:journals/vldb/VerwiebeGTM23}.

To program in \holon's procedural API, the programmer writes a single processing function.
This single processing function may combine various Windowed CRDTs, local values, and local windowed values.
It may also nest other processing functions, and branch on values.
Whereas the programmer is responsible for processing the inputs and producing outputs, the underlying runtime system will take care of the automatic synchronization of the shared state represented by the Windowed CRDTs, as well as the checkpointing and recovery.

\subsection{Guarantees}

The underlying execution engine manages the state of the implemented processing function, as well as its inputs and outputs.
It guarantees \emph{exactly-once processing}.
That is, for each partition, it guarantees that each event is processed exactly once, and that its side effects are reflected exactly once in the state, whereas outputs may be duplicated.
Even though the outputs may be duplicated, we consider them exactly-once, as they can be deduplicated by a consumer maintaining a map from partitions to window numbers by tagging the outputs by partition and window number.
Furthermore, the input is guaranteed to be consumed in a \emph{deterministic order per partition}, even after failures.

Windowed CRDTs are guaranteed to have global determinism.
That is, \emph{if a getWindowValue(ts) operation completes, then it is guaranteed to always return the same value for the same timestamp ts on every partition.}
This applies even though the WCRDT may be synchronizing its state in the background in a nondeterministic order.

\emph{The procedural API is guaranteed to be deterministic under the condition that it reads window values by the safe mode.}
If the unsafe mode is used, then the program is only guaranteed to be deterministic if it is implemented properly.
\autoref{lst:query-proc} shows an example of a safe use of the unsafe mode.
Notice how the listing semantically produces the same output with the unsafe mode as it would with the safe mode.
This is because this particular listing is structured such that data dependencies are acyclic and the window values are processed in sequence, thus the nondeterministic timing of when a window value is completed and available does not influence the result.
Programs in the dataflow API are of this structure, hence why they are always deterministic.
The difference between the different modes and APIs, is that the dataflow API allows for a pipelined execution, something which is possible to implement manually in the procedural API using the unsafe mode.
In the procedural API with the safe mode, reads of the window values block until the window is completed; a pipelined execution which avoids blocking is still possible by passing a continuation to be called once the window is completed.

\section{System Overview}\label{sec:system-overview}
The \holon\ system uses decentralized coordination.
It does so by removing direct dependencies between any two executing nodes.
That is, nodes do not directly exchange events by shuffling, and there is no pipeline of nodes, such that no node depends on another.
Instead, eventually consistent shared replicated state is used as the only means of synchronization and communication between nodes.
In this section, we describe the system architecture and its components, as well as the main algorithms, leveraging Windowed CRDTs for exactly-once processing.

\subsection{Architecture}
\begin{figure}[t]
  \centering
  \includegraphics[width=1.0\linewidth]{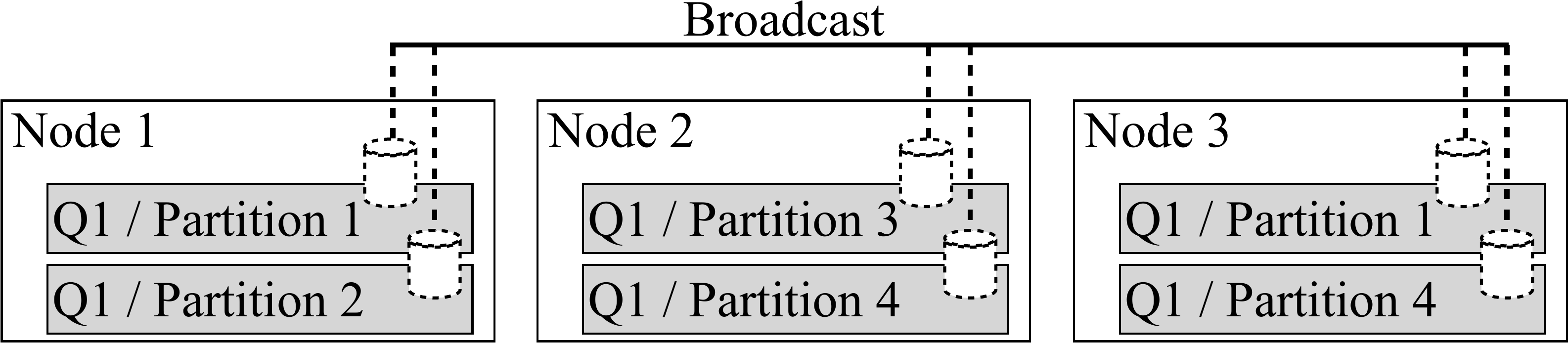}
  \caption{\holon\ deployment.}\label{fig:architecture-deployment}
\end{figure}

The system is composed of a set of decentralized nodes (\autoref{fig:architecture-deployment}).
Each node consumes events from a logged input stream, and produces events to a logged output stream.
Additionally, nodes may also consume from and produce events to a broadcast stream.
There may be an overlap in the processed partitions between nodes, for example, node 1 and node 3 may both process partition 1.
This is not an issue as the processing functions are deterministic, and the output is idempotent.

\begin{figure}[t]
  \centering
  \includegraphics[width=1.0\linewidth]{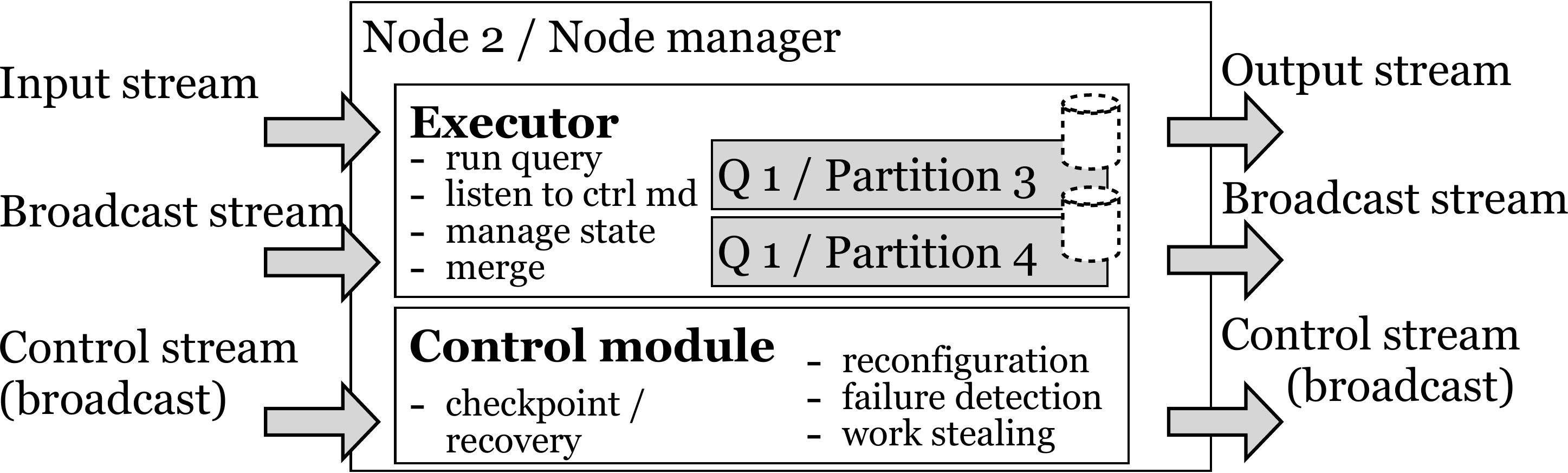}
  \caption{\holon\ node.}\label{fig:architecture-node}
\end{figure}

Inside of a node, the node processes multiple partitions in parallel (\autoref{fig:architecture-node}).
The executor processes events from the input stream, and the control module listens to control events, and informs the executor of any changes, such as reconfiguration instructions, detected failures, new work, etc.

\subsection{Windowed CRDTs}
Windowed CRDTs take a regular CRDT as input in the constructor, and produce a windowed~\cite{DBLP:conf/sigmod/LiMTPT05,DBLP:journals/pvldb/AkidauBCCFLMMPS15,DBLP:journals/vldb/VerwiebeGTM23} version of it (\autoref{alg:wcrdts}).

\begin{algorithm}[t]
  \caption{Windowed CRDTs.}\label{alg:wcrdts}
  \begin{algorithmic}[1]
    \State \textbf{type} \textsc{CRDT} \textbf{with default} zero
    \State \textbf{state} $windows$: Map \{ \textit{TimeStamp} $\mapsto$ \textsc{CRDT} \}
    \State \textbf{state} $progress$: Map \{ \textit{Node} $\mapsto$ \textit{TimeStamp} \}

    \Statex
    \Procedure{Insert}{element, ts}
    \If{$ts < progress[self]$} \textbf{error} \EndIf
    \State $windows[\text{windowOf}(ts)].\Call{insert}{element, ts}$
    \EndProcedure
    \Statex

    \Procedure{WindowValue}{ts}
    \If{$\text{window}(\Call{GlobalWatermark}{~}) < ts$}
    \State \Return $\textsc{None}$
    \EndIf
    \State \Return $\textsc{Some}(windows[ts].\Call{value}{})$
    \EndProcedure
    \Statex

    \Procedure{IncrementWatermark}{ts}
    \If{$progress[self] < ts$}
    \State $progress[self] \gets ts$
    \EndIf
    \EndProcedure
    \Statex

    \Procedure{GlobalWatermark}{~}
    \State \Return $\min(progress)$
    \EndProcedure
    \Statex

    \Procedure{Merge}{other}
    \ForAll{$(w,\,win)$ \textbf{in} $other.windows$}
    \State $windows[w] \gets windows[w].\Call{merge}{win}$
    \EndFor
    \ForAll{$(n,\,ts)$ \textbf{in} $other.progress$}
    \If{$progress[n] < ts$}
    \State $progress[n] \gets ts$
    \EndIf
    \EndFor
    \EndProcedure
  \end{algorithmic}
\end{algorithm}

The algorithm is a light-weight wrapper around CRDTs.
An event is simply inserted into the CRDT for the corresponding window of the event's timestamp.
Additionally, the algorithm uses a map of progress timestamps, representing the local watermark of each node, \ie the lowest timestamp of events that it may still process.
Using this, reading the window value is only possible if the global watermark is larger than the timestamp of the window.
Once this is the case, it is known that no more events for the window will be processed.
Thus, the window is completed.
Consequently, the window value is final, and will be the same for all nodes.

As discussed in the previous section, WCRDTs provide
\emph{strong eventual consistency},
\emph{global determinism of WCRDTs}, and
\emph{global determinism of programs with WCRDTs}.

Strong eventual consistency follows automatically from the use of the underlying CRDTs, as the WCRDTs state forms a lattice.
Global determinism of WCRDTs follows from the strong eventual consistency of the underlying CRDTs, and the use of progress watermarks.
Once all windows and their progress have merged, and the global watermark is larger than the window's timestamp, then it is guaranteed that there will be no more updates to the window.
Thus, once they have converged, they will remain the same.
Global determinism of programs with WCRDTs follows from the following observation.
Only stable values are read from the WCRDTs.
That is, if a (non-\lstinline{None}) value is returned by the method \lstinline{getWindowValue}, then the global watermark must be larger than the timestamp of the window.
This implies that no more updates to this window will be made.
Additionally, it also implies that all updates that \emph{have} been made to this window are visible and merged.
This way, even under nondeterministic order of events across separate nodes, and even if there exist cyclic dependencies between data, only stable values are read, which then inductively only depend on stable values themselves.
Thus, all read values are stable and so also deterministic.

\subsection{Failure Recovery}

The algorithm (\autoref{alg:failure-recovery}) executes a piece of work for a partition from some index on the input stream on that partition.
Sometimes (\ie read as ``at regular intervals''), it may store the partition state back to the storage.
It may also choose to steal work from another node, by invoking the recovery procedure on the stolen partitions.
The recovery procedure after failure, and reconfiguration, is not shown, as it follows a similar pattern.
The partition state itself forms a CRDT and is in turn broadcast to all nodes.
The lattice merge of a particular partition-\lstinline{id} is done by keeping the state with the largest \lstinline{nxtIdx}.

\algblockdefx{Sometimes}{EndSometimes}{\textbf{sometimes do}}{}
\algtext*{EndSometimes}

\begin{algorithm}[t]
  \caption{Failure recovery\label{alg:failure-recovery}}
  \begin{algorithmic}[1]
    \State \textbf{state} $partitions$: Map \{ \textit{PartitionId} $\mapsto$ \textit{PartitionState} \}
    \Statex

    \Procedure{Recover}{partitionId}
    \If{$partitions.\Call{contains}{partitionId}$}
    \Return
    \EndIf
    \State $partitions[partitionId] \gets storage.\Call{get}{partitionId}$
    \EndProcedure
    \Statex

    \Procedure{Run}{~}
    \While{true}
    \Sometimes
    \State $(id, (idx, odx, state)) \gets \Call{random}{partitions}$
    \State $(input,\;nxtIdx) \gets inStream.\Call{read}{id,\;idx}$
    \State $(output, nxtState) \gets \Call{run\_batch}{state,\;input}$
    \State $outStream.\Call{write}{id, odx, output}$
    \State $nxtOdx \gets odx + \Call{size}{output}$
    \State $partitions[id] \gets (nxtIdx, nxtOdx, nxtState)$
    \EndSometimes
    \Sometimes
    \State storage.\Call{put}{p,\;partitions[p]}%
    \EndSometimes%
    \Sometimes
    \State $ids \gets \Call{WorkSteal}{~}$
    \ForAll{$id \in ids$}
    \State \Call{Recover}{id}
    \EndFor
    \EndSometimes
    \EndWhile
    \EndProcedure
  \end{algorithmic}
\end{algorithm}

The failure recovery algorithm is relatively short.
This is an outcome of the determinism in the programming model.
The execution allows multiple nodes to process the same partitions; this will not lead to any issues, as they are deterministic and the output is idempotent.
For this reason, we can decouple the failure detection, reconfiguration, and work distribution from the execution, potential failure recovery, and reconfiguration.

\subsection{Implementation}
The implementation of \holon~ is available on GitHub licensed under Apache 2.0.\footnote{\url{https://github.com/portals-project/holon-streaming}}
The backend uses Apache Kafka topics for its logged input and output streams.
The broadcast and control streams are also implemented as Kafka topics.
The Windowed CRDTs are implemented as wrappers around Akka/Pekko Distributed Data's~\cite{akkadd,pekkodd} CRDTs for representing the underlying CRDTs.
The nodes are run on Google Cloud using Docker containers and Kubernetes for orchestration.
The current implementation of \holon\ is limited to tumbling windows and partition-ordered streams, but we plan to support other window types and out-of-order streams in the future.

\section{Evaluation}\label{sec:evaluation}
The evaluation is divided into two parts.
The failure recovery and reconfiguration experiments evaluate the system's performance on a distributed deployment with injected failures, comparing its performance to Apache Flink~\cite{DBLP:journals/debu/CarboneKEMHT15}.
The scalability experiments evaluate the system's performance when scaling up the number of nodes.

\subsection{Experimental Setup}
The end-to-end benchmarks are run on a distributed deployment on Google Cloud.
For the \holon\ experiments, we have used
2vCPU nodes
(GCP, 4GB memory per node), with up to 100 nodes, where 20 nodes are used for Kafka, 20 nodes for producing and consuming Nexmark events, and 60 nodes for the \holon\ system.
The number of deployed execution nodes is described under each experiment.

The experiments run a set of workloads, which all represent different types of global aggregations.
For this, we are running global aggregation queries from the Nexmark benchmark~\cite{tucker2002nexmark}.
\begin{itemize}
  \item Nexmark Q0: Pass through.
  \item Nexmark Q4: Average price per category.
  \item Nexmark Q7: Highest bids.
\end{itemize}

Where Q0 is a stateless query, Q4 requires either a shuffle of the data by category together with category-local aggregations, or a global aggregation by category without shuffles, and Q7 requires a global aggregation for calculating the globally highest bids.
The following metrics are collected when evaluating the workloads.
\begin{itemize}
  \item Throughput: Measured as the total number of consumed events per second.
  \item Latency: End-to-end latency measured by Kafka insertion timestamps.
\end{itemize}
Additionally, we calculate these derived metrics from the above.
\begin{itemize}
  \item Sensitivity~\cite{sensitivity2025}: The measured deviation caused by failures against the failure-free expectation. For example, it may measure both the amplitute and duration of the failure's effect on latency.
\end{itemize}

As a baseline, we have chosen to use Apache Flink~\cite{DBLP:journals/debu/CarboneKEMHT15} as a representative of a state-of-the-art production stream processing system. Flink has been shown to have good performance and fast failure recovery in comparison to related systems~\cite{DBLP:conf/debs/VogelHPER24}.
We used Flink v2.0, deployed on GCP Kubernetes Engine, with RocksDB state backend (incremental, unaligned checkpointing with 5 second interval and mode EXACTLY\_ONCE),
and heartbeat configuration set to lower values than the recommended Flink defaults~\cite{DBLP:conf/sigmod/SilvestreFSK21}: interval of 4 seconds (default: 10 seconds) and a timeout of 6 seconds (default: 60 seconds).
Each TaskManager has 8GB allocated to avoid out-of-memory issues, with a default network buffer allocation of 10\%.

\begin{figure*}[t]
  \centering
  \begin{subfigure}[c]{0.9\textwidth}
    \setlength{\abovecaptionskip}{0pt}
    \includegraphics[width=\linewidth]{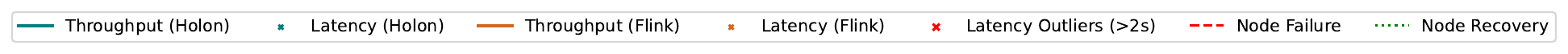}
  \end{subfigure}
  \vspace{0em}
  \begin{subfigure}[c]{0.33\textwidth}
    \setlength{\abovecaptionskip}{0pt}
    \includegraphics[width=\linewidth]{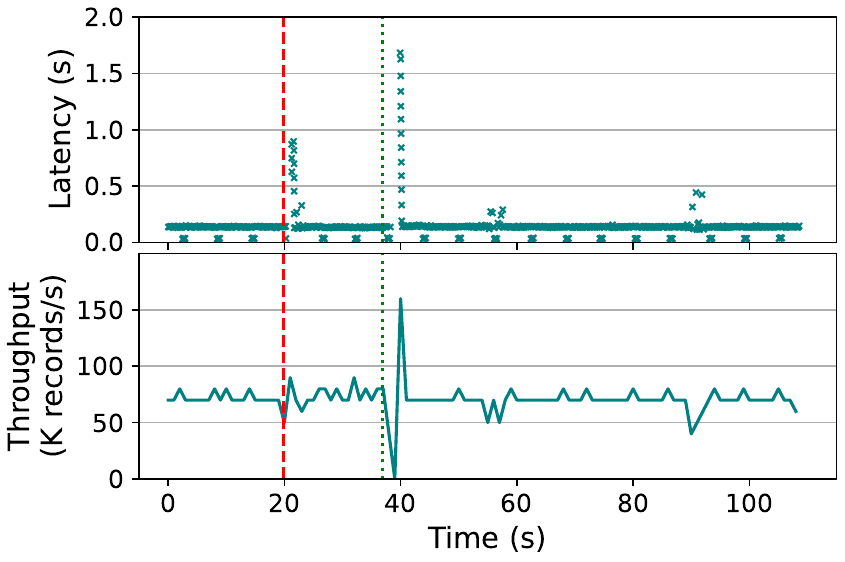}
  \end{subfigure}
  \begin{subfigure}[c]{0.33\textwidth}
    \setlength{\abovecaptionskip}{0pt}
    \includegraphics[width=\linewidth]{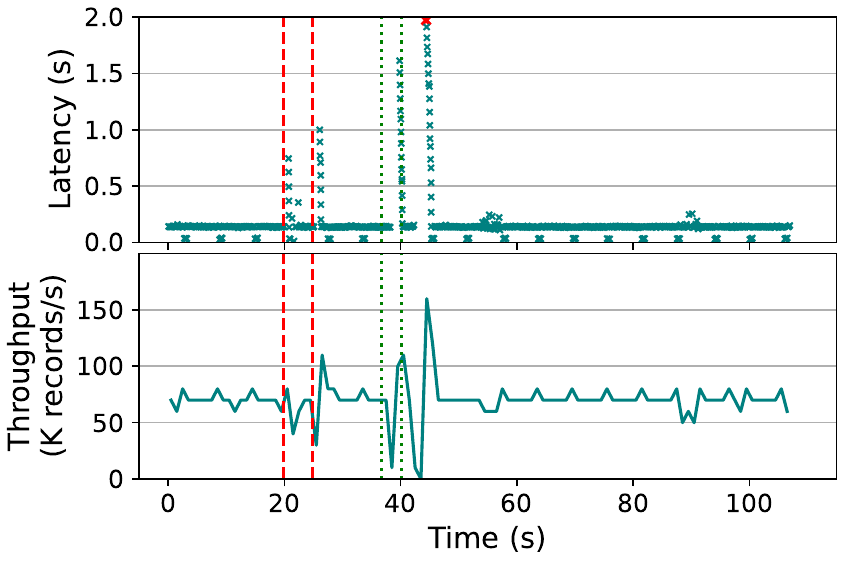}
  \end{subfigure}
  \begin{subfigure}[c]{0.33\textwidth}
    \setlength{\abovecaptionskip}{0pt}
    \includegraphics[width=\linewidth]{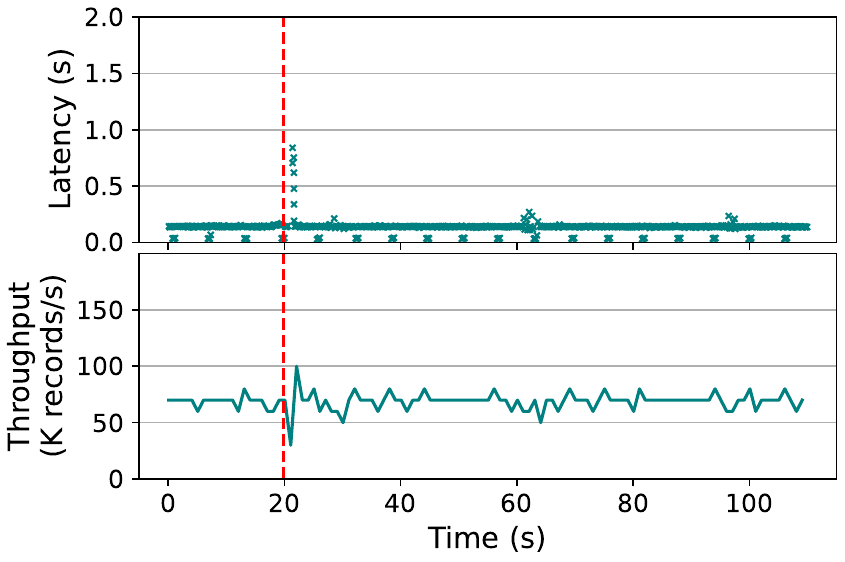}
  \end{subfigure}

  \vspace{-1.1em}

  \begin{subfigure}[c]{0.33\textwidth}
    \setlength{\abovecaptionskip}{0pt}
    \includegraphics[width=\linewidth]{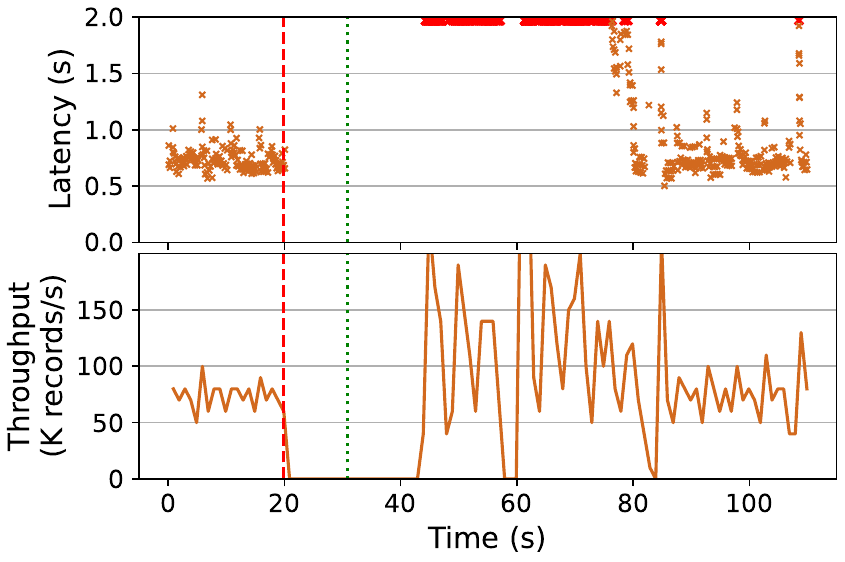}
    \caption{Concurrent Failures}
  \end{subfigure}
  \begin{subfigure}[c]{0.33\textwidth}
    \setlength{\abovecaptionskip}{0pt}
    \includegraphics[width=\linewidth]{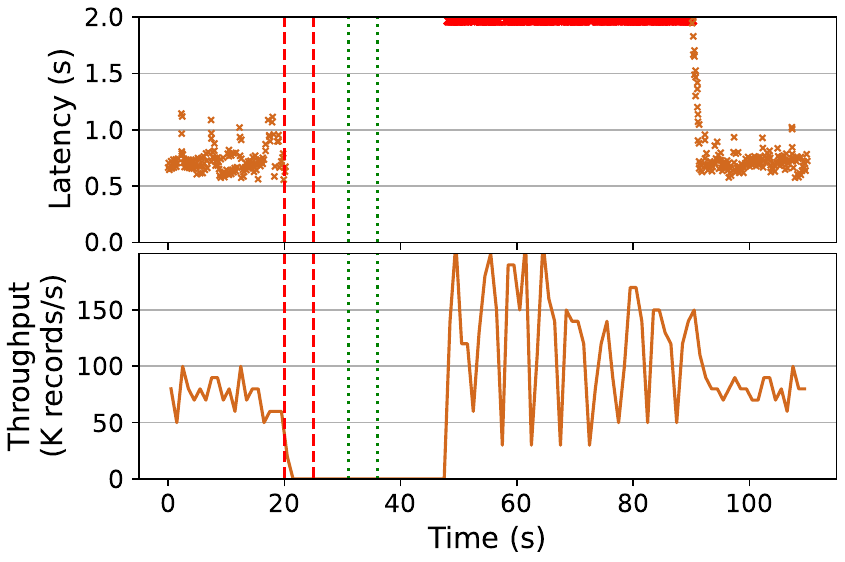}
    \caption{Subsequent Failures}
  \end{subfigure}
  \begin{subfigure}[c]{0.33\textwidth}
    \setlength{\abovecaptionskip}{0pt}
    \includegraphics[width=\linewidth]{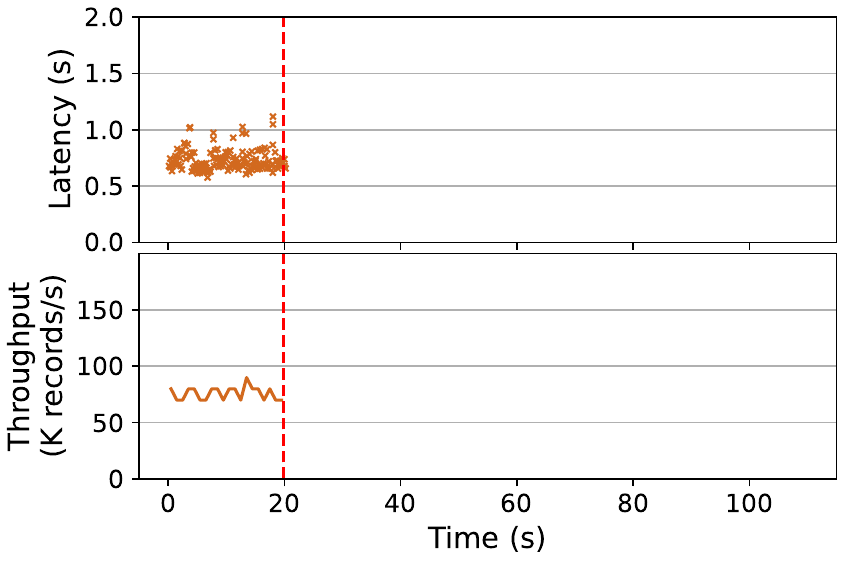}
    \caption{Crash Failures}
  \end{subfigure}
  \vspace{-0.7em}
  \caption{Latency \& throughput during node failure scenarios (Holon top - Flink bottom)}
  \label{fig:failure-recovery}
\end{figure*}

\begin{figure}[t]
  \setlength{\abovecaptionskip}{0pt}
  \centering
  \includegraphics[width=1.0\linewidth]{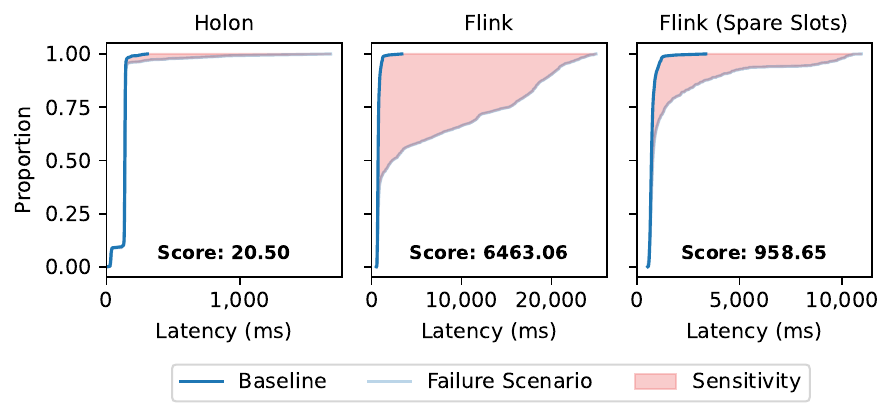}
  \caption{Latency sensitivity curves for concurrent failures.}\label{fig:latency-sensitivity-curves}
\end{figure}

\begin{figure}[t]
  \setlength{\abovecaptionskip}{0pt}
  \centering
  \includegraphics[width=1.0\linewidth]{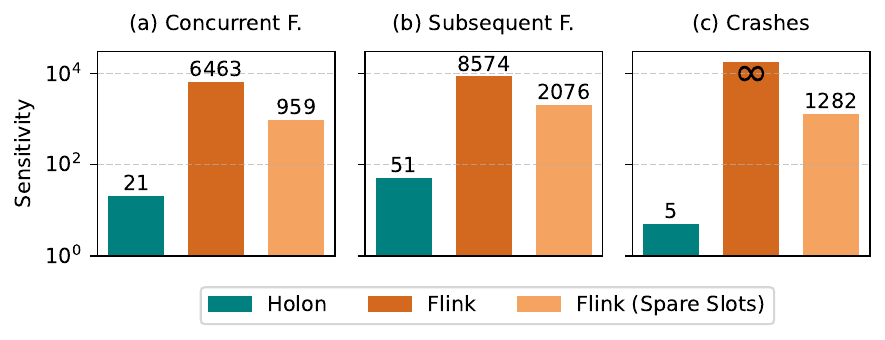}
  \caption{Latency sensitivity across failure scenarios.}\label{fig:failure-recovery-sensitivity}
\end{figure}

\subsection{Failure Recovery and Reconfiguration}

To measure the impact of failures and failure recovery on throughput and latency, we run workload Q7 on a deployment of five nodes, and inject failures by manually stopping and restarting nodes.
We consider three failure scenarios:
\begin{itemize}
  \item Concurrent node failures. Two nodes are failed at the same time, and restarted 10 seconds later.
  \item Subsequent node failures. Two nodes are failed within 5 seconds of each other, and restarted 10 seconds after failure.
  \item Crash failures. Two nodes are failed and not restarted.
\end{itemize}

\autoref{fig:failure-recovery} shows the end-to-end latency of processed windows and record level throughput over time.
As can be seen, Holon is able to fully recover within 2 seconds after a failure, catching up with lost processing time.
Flink, on the other hand, has a much longer recovery time, taking up to 70 seconds to fully recover.
Furthermore, as can be seen in the case of a crash, whereas \holon\ continues processing after automatic reconfiguration, Flink will stop processing in the case that its slots are full.
Our measured recovery times for Flink (35-70 seconds) are in close agreement with those reported by Vogel et al.~\cite{DBLP:conf/debs/VogelHPER24}, who observed 40-60 seconds recovery under comparable failure scenarios.

To quantify the impact on latency, we calculate the latency sensitivity due to the failures as shown in \autoref{fig:failure-recovery-sensitivity}.
It shows that \holon\ has a factor 20 or more lower sensitivity than Flink.
This is due to the much faster recovery time because of the work-stealing mechanism and its deterministic programming model.
The latency sensitivity curves are shown in \autoref{fig:latency-sensitivity-curves}.
The sensitivity is the area between the curve with failures and the baseline without failures.

This is reinforced by the P99 numbers in \autoref{tab:failure-recovery-latency-comparison}.
For the baseline without failures, \holon\ has an average end-to-end latency of 0.13 seconds, whereas Flink has an average latency of 0.77 seconds.
For the p99 latency, \holon\ has 0.19 seconds, whereas Flink has 1.74 seconds.
For the average latency under the baseline scenario in \autoref{tab:failure-recovery-latency-comparison}, \holon\ has a factor 5 lower latency than Flink, and under failure scenarios \holon\ has a factor 11 or more lower latency than Flink.

\subsection{Scalability}

\begin{figure}[t]
  \includegraphics[width=\linewidth]{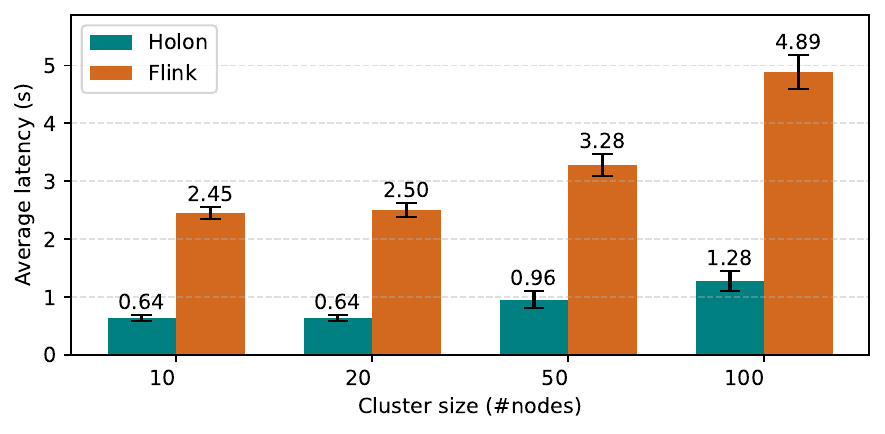}
  \caption{Average latency for Q7.}
  \label{fig:scalability-latency-bars}
\end{figure}

\autoref{fig:scalability-latency-bars} shows the latency for Q7 for cluster sizes from 10 to 100 nodes with a global ingestion rate of 10k events per second per node, where the maximum throughput is 1M events per second.
This experiment (\autoref{fig:scalability-latency-bars}) was measured on a single 64-core server (503 GB, 128 threads) with 100 producer nodes and 100 \holon\ nodes, scaling input volume with cluster size.
As can be seen, \holon\ achieves better latency for all cluster sizes.
For cluster size of 10 nodes, \holon\ has an average latency of 0.64 seconds, whereas Flink has an average latency of 2.45 seconds, a difference of factor 3.8.

This experiment was repeated to measure the maximum throughput achieved for 10 nodes with 50 partitions and 50 producers.
The ingestion rate starts at 1K events per second per producer and is increased exponentially over time.
The maximum throughput is measured as the peak total throughput of the system before it either drops or stabilizes.
\holon\ achieves 11x the throughput compared to Flink for Q4 (122k vs 10.4k events/s), and 1.8x for Q7 (2.05M vs 1.09M events/s).

\begin{table}[t]
  \centering
  \caption{Latency comparison (in seconds) under different failure scenarios.}
  \label{tab:failure-recovery-latency-comparison}
  \resizebox{\linewidth}{!}{
    \begin{tabular}{l
        *{2}{c}
        *{2}{c}
        *{2}{c}
        *{2}{c}
      }
      \toprule
      \textbf{System} &
      \multicolumn{2}{c}{\textbf{Baseline}} &
      \multicolumn{2}{c}{\makecell{\textbf{Concurrent} \\ \textbf{Failures}}} &
      \multicolumn{2}{c}{\makecell{\textbf{Subsequent} \\ \textbf{Failures}}} &
      \multicolumn{2}{c}{\makecell{\textbf{Crash} \\ \textbf{Failures}}} \\
      & Avg. & P99 & Avg. & P99 & Avg. & P99 & Avg. & P99 \\
      \midrule
      Holon & 0.13 & 0.19 & 0.15 & 0.88 & 0.18 & 1.54 & 0.13 & 0.23 \\
      Flink & 0.77 & 1.74 & 7.24 & 23.94 & 9.36 & 28.01 & -- & -- \\
      Flink (Spare Slots) & 0.77 & 1.74 & 1.72 & 10.24 & 2.85 & 13.88 & 2.05 & 7.95 \\
      \bottomrule
    \end{tabular}
  }
\end{table}

\section{Related Work}\label{sec:related-work}
To the best of our knowledge, \holon\ is the first exactly-once stream processing system which has integrated CRDTs.
The presented Windowed CRDTs are a novel data structure which unify windowing~\cite{DBLP:conf/sigmod/LiMTPT05,DBLP:journals/pvldb/AkidauBCCFLMMPS15,DBLP:journals/vldb/VerwiebeGTM23} with conflict-free replicated data types (CRDTs)~\cite{DBLP:conf/sss/ShapiroPBZ11,DBLP:journals/jpdc/AlmeidaSB18}.
A bounded LWW-set for computing aggregates over a bounded set (sliding window) of updates to a CRDT has been proposed by Meiklejohn et al.~\cite{DBLP:conf/ccnc/MeiklejohnHR16}, such as an aggregation on the latest $n$ updates. However, it is unclear if every replica outputs the same sequence of aggregates in~\cite{DBLP:conf/ccnc/MeiklejohnHR16}, as is done by the windowed CRDTs in this paper.
Adas and Friedman~\cite{DBLP:conf/srds/AdasF21} use CRDTs for sliding window sketches, which are approximate aggregations of the data, whereas our work is on exact aggregations.
Hydroflow~\cite{samuel2021hydroflow} and the Hydro project~\cite{DBLP:conf/applied/HellersteinLMPS23,cheung2021new} is a stream processing system which uses CRDTs for distributed computations, but it does not provide exactly-once processing guarantees.

Similar to \holon, existing work on CRDTs has focused on providing additional guarantees for their use.
This includes: sequential ordering on eventually consistent data~\cite{DBLP:conf/ecoop/BurckhardtLPF15};
methods for flushing local updates~\cite{DBLP:conf/ecoop/BurckhardtFLW12};
observable atomic consistency for operations on CRDTs~\cite{DBLP:journals/jlap/ZhaoH20};
verifiable guarantees such as invariant preservation~\cite{DBLP:journals/toplas/HaasMYBM24};
replicated protocols on top of CRDTs and similar data types for stronger guarantees~\cite{DBLP:journals/corr/abs-2405-15578,DBLP:journals/corr/abs-2504-05173};
causal stability for the purpose of metedata compaction among others~\cite{baquero2017pure,bauwens2020causality}; and deriving CRDTs from sequential data types~\cite{DBLP:conf/dais/PorreMTSMB19,DBLP:journals/pacmpl/PorreFPB21}.

Failure recovery~\cite{DBLP:journals/csur/ElnozahyAWJ02} is an integral aspect of stream processing systems with exactly-once processing guarantees~\cite{DBLP:journals/vldb/FragkoulisCKK24}.
Apache Flink takes asynchronous snapshots of the system state, which is committed by means of a centralized two-phase commit protocol~\cite{DBLP:journals/debu/CarboneKEMHT15}.
Other streaming systems have a centralized coordinator such as Google Dataflow~\cite{DBLP:journals/pvldb/AkidauBCCFLMMPS15}, or each processing node coordinates its own as in Apache Kafka Streams~\cite{kreps2011kafka,sax2018streams}.
Rhino~\cite{del2020rhino} provides on-the-fly reconfiguration support for state migration implemented on Apache Flink, making reconfigurations much faster.
Similarly, work by Burckhardt et al.~\cite{DBLP:journals/pvldb/BurckhardtCGJKM22} use a decentralized approach to failure recovery.
Leveraging determinism for failure recovery has been proposed for example by Silvestre et al.~\cite{DBLP:conf/sigmod/SilvestreFSK21} within the context of Apache Flink.
On-the-fly reconfiguration support for state migration for Apache Flink has been shown to make reconfigurations much faster~\cite{DBLP:conf/sigmod/MonteZRM20}.
Similarly, using remote storage as the primary storage for checkpoints with local caching has been shown to reduce recovery time significantly~\cite{DBLP:journals/pvldb/MeiLHLYXHCKW25}.
Similar to our execution model consisting of a single node executing the entire pipeline, this is also the approach taken in Naiad / Timely Dataflow~\cite{murray2016incremental,murray2013naiad}.

Global aggregations may either be done by tree reductions or by other means.
Watermarks help understanding disordered and unbounded streams~\cite{akidau2021watermarks}.
Traub et al.~\cite{traub2019efficient} demonstrated that window aggregations can leverage algebraic properties such as commutativity and associativity for improved performance, however, it was not applied to global aggregations as is done in this paper.
Apache Spark~\cite{zaharia2012resilient,zaharia2010spark} has a method \emph{treeReduce}~\cite{sparktreereduce} which performs a tree reduction of the data into a aggregate result.
In Apache Flink~\cite{DBLP:journals/debu/CarboneKEMHT15}, tree reduction can be done manually, by setting the parallelism level, and by creating reduction operators.
Window aggregations are supported in Apache Kafka's KSQL~\cite{DBLP:conf/edbt/JafarpourD19}.

\section{Conclusions and Future Work}\label{sec:conclusions}
We have presented \holon, a stream processing system with decentralized coordination for global aggregations.
It is based on a novel programming model which uses \emph{Windowed CRDTs} for shared replicated state.
The programming model is deterministic and uses decentralized cooridnation by leveraging the guarantees of Windowed CRDTs.
Thus, making global aggregations scalable and low-latency.
The results show that \holon~ achieves up to 5-times lower latency and 2-times higher throughput than Apache Flink for global aggregation queries from the Nexmark benchmark~\cite{tucker2002nexmark}.
The performance benefits are attenuated under failures, demonstrating an 11-times lower latency under failure scenarios.
As demonstrated, Windowed CRDTs are powerful abstractions for global aggregations in stream processing systems, with a familiar interface to existing users, providing great performance with easy-to-understand guarantees.

We are working on several extensions to the programming model and system for future work.
\emph{Delta-based CRDTs}~\cite{DBLP:journals/jpdc/AlmeidaSB18}
differ to regular CRDTs by producing incremental synchronization updates.
They naturally integrate into the streaming-based programming model, as the output of a delta-based WCRDT is a stream of deltas.
By using delta-based WCRDTs, it would be possible to incrementally synchronize large states, thus incrementalizing the state synchronization.
\emph{Data-based windows}
are windows whose boundaries are defined by the underlying data itself, which is an interesting opportunity for generalizing the windowing abstraction.
\emph{Multi-threading}
within the context of local processing functions is naturally supported by the use of CRDTs.
A sequence of events can be handled in parallel, enabling scaling-up execution nodes to a large number of cores.

\begin{acks}
  We would like to thank Wu Tianxing for significant contributions to a concurrent draft of a companion paper with title ``An Operational Semantics for Windowed Computations on CRDTs''.

  This study was partially funded by Digital Futures under a Research Pairs Consolidator grant (PORTALS).
\end{acks}

\bibliographystyle{ACM-Reference-Format}
\bibliography{refs}

% ...

\end{document}